\documentclass{aaold}
\usepackage{amssymb}

\usepackage{graphicx}

\begin{document}

\title{Hemispherical transport equation: modeling of quasiparallel collisionless shocks}
\author{V.N. Zirakashvili \inst{1,2}}
\date{Received ; accepted }
\titlerunning{Hemispherical transport equation}

%\fnmsep\thanks{%
%Just to show the usage of the elements in the author field}}

\subtitle{}

\offprints{V.N. Zirakashvili, zirak@izmiran.ru}

\institute{Pushkov Institute of Terrestrial Magnetism, Ionosphere and Radiowave
Propagation, 142190, Troitsk, Moscow Region, Russia
%             \email{Zirak@izmiran.rssi.ru}
\and
Max-Planck-Institut f\"{u}r\ Kernphysik, D-69029,
Heidelberg, Postfach 103980, Germany
%              \email{Heinrich.Voelk@mpi-hd.mpg.de}
}
\abstract{ Using a so-called hemispherical model we
derive a general transport equation for cosmic ray and thermal particles scattered
in pitch angle by magnetic inhomogeneities in a moving collisionless plasma.
The weak scattering through 90 degrees results in
isotropic particle distributions in each hemisphere. The consideration is not limited
by small anisotropies and by the condition that
particle velocities are higher than characteristic flow velocity differences.
For high velocities and small
anisotropies the standard
cosmic ray transport equation is recovered. We apply the equations derived for
investigation of injection and acceleration of particles by collisionless shocks.

}

%________________________________________________________________
\maketitle

\section{Introduction}

Progress in cosmic ray astrophysics has been made with the introduction of the transport
equation for cosmic rays (Krymsky \cite{krymsky64}, Parker \cite{parker65},
Dolginov \& Toptygin \cite{dolginov66}, Gleeson \& Axford \cite{gleeson69}). It was
successfully applied to many astrophysical problems: cosmic ray modulation
in the solar wind, acceleration of particles at shock fronts, cosmic ray propagation
in the Galaxy etc. This equation implies that in the strong scattering approximation
the particle distribution
is almost isotropic, which in particular implies that  velocities of particles are higher
than the characteristic flow velocity differences in the plasma.
However, in several circumstances these conditions are violated and more general kinetic equations
should be used which also deal with the evolution of the pitch angle distribution.
Since the solution of such equations is not a simple task,
we would like find a description that comes close to that implied by the standard
transport equation.

Energetic particles are scattered in pitch angle relative to the mean magnetic field direction
by random magnetic inhomogeneities. It is well known that
in the case of particle transport along the mean magnetic field a so-called problem of
scattering through the pitch-angle of 90 degrees exist. According to quasilinear theory the
resonant scattering of particles is weak at small pitch-angle cosines
(Jokipii \cite{jokipii66}). The main reason for this phenomenon is the small energy density
of short waves propagating along the magnetic field and scattering the particles 
with pitch angles close
to 90 degrees. These waves may be easily damped by thermal ions. Even without such a damping,
the scattering in the vicinity of 90 degrees is depressed in the case of a    power-law spectrum
of waves that is proportional to $k^{-2}$ or steeper. Here $k$ is the wavenumber. This problem
can be avoided if one takes into account the scattering by oblique waves or nonlinear interactions,
for example magnetic mirroring (see e.g. V\"olk \cite{voelk73}, \cite{voelk75}; Achterberg
\cite{achterberg81}). One can expect that the scattering efficiency is small near the
pitch angle of 90 degrees.

Observations
of suprathermal pick-up particles in the solar wind suggest that this is the case
(Fisk et al. \cite{fisk97}). If the
scattering through 90 degrees is weak, the particle distribution is almost isotropic
 in each hemisphere of the velocity directions parallel or antiparallel to the mean field.
 So, in this case the angular dependence of particle distribution
is reduced to the two number densities of forward and backward moving particles. The corresponding
equations were derived by Isenberg (\cite{isenberg97}) for the case of pick-up ions in the
solar wind.

We use this approach and derive general transport equations for arbitrary nonrelativistic
flow velocities. We use the particle distributions in the frame moving with the
medium. This permits us to consider the case of arbitrary particle velocities.
Since the number density of particles in both hemispheres may substantially differ,
large anisotropies can also be taken into account. The equations derived have a broad
range of applicability: propagation of solar energetic particles and pick-up ions in the
solar wind, injection into diffusive shock acceleration and other processes.

As an example of the application of the equations derived, we consider the problem of
diffusive shock acceleration. Since
our equations describe propagation of thermal as well as energetic particles in the vicinity
of the collisionless shock front, they can be used for
the determination of the shock velocity profile and the spectra of thermal and nonthermal
particles without any assumptions regarding the injection efficiency. One only has to
prescribe the scattering law. In this sense the approach is
similar to the one used by Ellison and Eichler
(\cite{ellison84}). They applied a Monte-Carlo technique for the solution of the kinetic
 equation.

  The basis of such an approach is the idea that the interaction of upstream and downstream
 plasmas may result in instabilities similar to firehose instability
 (Parker \cite{parker61}, Quest \cite{quest88}). The magnetic inhomogeneities produced by
 such instabilities may play the role of scattering centers and provide the shock
 dissipation and heating. These ideas are also the basis of the so-called hybrid simulations of
 collisionless shocks (Leroy \& Winske \cite{leroy83}, Quest \cite{quest86}). The ions move in
 self-consistent electromagnetic fields and the electrons are considered as a charged neutralizing
 fluid in these simulations.

The equations are derived in the next two Sections. Application to the case
of collisionless shock is considered
in Sect.4 and 5. Sect.6 contains
a discussion of results obtained and conclusions.

\section{Basic equations}
We start with the kinetic equation for the momentum distribution
$f({\bf r},{\bf p})$ of charged particles
\begin{equation}
\frac {\partial f}{\partial t}+{\bf v}\nabla f+
{\bf F}_L\frac {\partial f}{\partial {\bf p}}=\hat {St} f
\end{equation}
Here ${\bf v}$ and  ${\bf p}$ are the particle velocity and momentum respectively,
${\bf F}_L={q\bf E}+(q/c)[{\bf v\times B}]$ is the Lorence force
and the operator $\hat {St}$ describes the pitch angle scattering of particles
by random magnetic inhomogeneities in the fluid frame. We shall further assume that the mean
electric field $\bf E$ can be
described as

\begin{equation}
{\bf E}=E_\parallel {\bf b}-\frac 1c[{\bf u}\times {\bf B}],
\end{equation}
where $\bf b$ is the unit vector along the direction of the mean magnetic field $\bf B$, $\bf u$
is the mass velocity and $E_\parallel $ is the component of the mean electric field which is parallel
to the mean magnetic filed.

It is convenient to perform a change of variable $\bf p$ to ${\bf p}'$ that formally
coincides with the nonrelativistic Lorence transform

\begin{equation}
{\bf p}'={\bf p}-\frac {{\bf u}}{v}p.
\end{equation}
Averaging the particle
gyrophase angle and neglecting terms of the order $u/c<<1$
we find (cf. Skilling \cite{skilling71},
Kulsrud \cite{kulsrud83}):

\[
\frac {\partial f}{\partial t}+(v'\mu '{\bf b}+{\bf u})\nabla f=\hat {St} f+
\]
\[
\frac {\partial f}{\partial p'}\left[
\frac {3\mu '^2-1}{2}p'
{\bf b(b\nabla )u}+\frac {1-\mu '^2}{2}p'\nabla {\bf u}-\mu 'F_\parallel
\right] +
\]
\begin{equation}
\frac {\partial f}{\partial \mu '}(1-\mu '^2)\left[
\frac 32\mu '{\bf b(b\nabla )u}-\frac {v'}{2}\nabla {\bf b}-
\frac {\mu '}{2}\nabla {\bf u}-\frac {F_\parallel }{p'}
\right]
\end{equation}
Here $\mu '$ is the cosine of the pitch angle of the particle and $F_\parallel $ is the
sum of the parallel components of the electric and inertia force:

\begin{equation}
F_\parallel =qE_\parallel -
{\bf b}\frac {p'}{v'}\left( \frac {\partial {\bf u}}{\partial t}+
{\bf (u\nabla )u}\right) .
\end{equation}
The scattering operator can be written as

\begin{equation}
\hat {St} f=\frac {\partial }{\partial \mu '}\frac {1-\mu '^2}{2}\nu (p',\mu ')
\frac {\partial f}{\partial \mu '}
\end{equation}
Here $\nu $ is the scattering frequency.

Let us assume that the scattering frequency is large for pitch angle cosines $|\mu '|>\mu _0$.
The momentum distribution $f$ is then isotropic in these regions of the angle space.
We introduce distributions of particles in the backward and forward hemispheres $N_\pm $:
$f(p',\mu ')=N_-(p')$ for $\mu '<-\mu _0$ and $f(p',\mu ')=N_+(p')$ for $\mu '>\mu _0$.
Integration of
Eq. (4) from -1 to $-\mu _0$ and from $\mu _0$ to $1$ gives the equations for these distributions:
%in the backward and the forward hemispheres $N_-(p')$ and $N_+(p')$ respectively:

\[
\frac {\partial N_\pm }{\partial t}+
\left( {\bf u}\pm \frac {v'}{2}{\bf b}\right) \nabla N_\pm
+ \left( \pm \frac 12F_\parallel -\frac {p'}{3}{\bf \nabla u}\right)
\frac {\partial N_\pm }{\partial p'}=
\]
\begin{equation}
\mp \left. \frac {\nu }{2}\frac {\partial f}{\partial \mu '}\right| _{\mu '=\pm \mu _0}
\end{equation}
It was assumed that $\mu _0<<1$.
Now we calculate the right-hand side of these equations. For this we must find the solution
of Eq. (4) in the region $|\mu '|<\mu _0$. Since $\mu _0<<1$, we can leave only the
terms with derivatives on $\mu '$ in this equation with the result:

\begin{equation}
-\left( \frac {v'}{2}\nabla {\bf b}+\frac {F_\parallel }{p'}\right)
\frac {\partial f}{\partial \mu '}=
\frac {\partial }{\partial \mu '}\frac {\nu}{2}
\frac {\partial f}{\partial \mu '}.
\end{equation}
This equation should be solved with the boundary conditions $f(p',\pm \mu _0)=N_\pm (p')$.
Substitution of the solution into the Eq. (7) then gives

\[
\frac {\partial N_\pm }{\partial t}+
\left( {\bf u}\pm \frac {v'}{2}{\bf b}\right) \nabla N_\pm
+ \left( \pm \frac 12F_\parallel -\frac {p'}{3}{\bf \nabla u}\right)
\frac {\partial N_\pm }{\partial p'}=
\]
\begin{equation}
\mp \nu _\mp (N_+-N_-).
\end{equation}
Here the frequencies $\nu _\pm $ describe the rate of particle exchange between forward and
backward hemispheres:

\begin{equation}
\nu _\pm (p')=\pm \frac {\frac {F_\parallel }{p'}+\frac {v'}{2}\nabla {\bf b}}
{\exp \left[ \pm \lambda \left(
\frac {2F_\parallel }{v'p'}+\nabla {\bf b}\right) \right] -1},
\end{equation}
and the mean free path $\lambda $ of the particle in the uniform medium
is given by the expression:

\begin{equation}
\lambda =v'\int ^{\mu _0}_{-\mu _0}\frac {d\mu '}{\nu (p',\mu ')}.
\end{equation}

A very similar equation was derived by Isenberg (\cite{isenberg97}) for pick-up ions
in the solar wind. We found more accurate expressions for the frequencies $\nu _\pm $
and these coincide
with the result of Kota (\cite{kota00}) obtained for the case $F_\parallel =0$.

 We neglected the derivatives in time and space in Eq. (8). This assumption
is valid if the characteristic time $\tau $ and the characteristic scale $l$ of the
problem considered are large enough: $\tau , \ l/v'>>\mu _0^2/\nu \sim \mu _0\lambda /v'$.
Since $\mu _0<<1$, our derivation is valid even if the scale $l$ is comparable with the
mean free path $\lambda $. The equation (9) is exact in the mathematical limit
$\mu _0\to 0$, $\lambda \to const$.

The Eqs. (9) can be rewritten in the conservative form:

\[
\frac {\partial N_\pm }{\partial t}+
\nabla \left( {\bf u}\pm \frac {v'}{2}{\bf b}\right) N_\pm
+
\]
\begin{equation}
\frac {1}{p'^2}\frac {\partial }{\partial p'}p'^2
\left( \pm \frac 12F_\parallel -\frac {p'}{3}{\bf \nabla u}\right)
N_\pm =
\nu _\mp N_\mp -\nu _\pm N_\pm .
\end{equation}

The equations derived have a simple physical meaning. The second term in the
left hand side describes the transport of the particles by
the medium moving with velocity ${\bf u}$ and the proper motion of
particles along the magnetic field with the speed $\pm v'/2$ (to be
compared with ``coherent'' propagation of solar energetic
particles considered by Earl (\cite{earl74})). The third term describes
the energy losses or gains in the inhomogeneous flow and electric field. The
right hand side corresponds to the exchange of particles between forward and backward
hemispheres.

It is clear that the system of equations (12) is hyperbolic. It
can be reduced to uncoupled equations for $N_+$ and $N_-$:

\[
\frac {\partial N_\pm }{\partial t}+{\bf u}\nabla N_\pm-
\frac {p'}3\frac {\partial N_\pm }{\partial p'}{\bf \nabla u}=
-\left[ \frac {\partial }{\partial t}+
\nabla \left( {\bf u}\mp \frac {v'}{2}{\bf b}\right) + \right.
\]
\[
\left. +
\frac {1}{p'^2}\frac {\partial }{\partial p'}p'^2
\left( \mp \frac 12F_\parallel -\frac {p'}{3}{\bf \nabla u}\right)
\right] \frac {1}{2\nu _\mp}\times
\]
\begin{equation}
\left[ \frac {\partial N_\pm }{\partial t}+
\left( {\bf u}\pm \frac {v'}{2}{\bf b}\right) \nabla N_\pm
+ \left( \pm \frac 12F_\parallel -\frac {p'}{3}{\bf \nabla u}\right)
\frac {\partial N_\pm }{\partial p'}\right] .
\end{equation}

In the case of high scattering frequencies the particle distributions in
the different hemispheres are almost equal to each other, $N_+\approx N_-\approx N$.
Assuming also a slow
time evolution and particle velocities much larger than the medium velocity we
come to the standard transport equation for cosmic rays (Krymsky \cite{krymsky64},
Parker \cite{parker65},
Dolginov \& Toptygin \cite{dolginov66}, Gleeson \& Axford \cite{gleeson69}):

\begin{equation}
\frac {\partial N}{\partial t}+{\bf u}\nabla N-
\frac {p'}3\frac {\partial N}{\partial p'}{\bf \nabla u}=
(\nabla {\bf b})D_\parallel ({\bf b}\nabla )N .
\end{equation}
Here the diffusion coefficient along the magnetic field $D_\parallel =v'\lambda /4$.

Another interesting case is when the energy changes and transport by
the medium are negligible. Then Eq. (13) is reduced to the so-called
telegraph equation (cf. Fisk \& Axford \cite{fisk69}):

\begin{equation}
\frac {\partial N_\pm }{\partial t}=
-\left[ \frac {\partial }{\partial t}
\mp \frac {v'}{2}(\nabla {\bf b}) \right]
\frac {1}{2\nu _\mp}\left[ \frac {\partial N_\pm }{\partial t}
\pm \frac {v'}{2}({\bf b}\nabla )N_\pm \right] .
\end{equation}

This equation describes propagation of cosmic ray particles with finite
velocity and can be reduced to the diffusion equation only in the case of a slow time
evolution.

\section{Self-consistent closure}

The equations derived in the previous section contain prescribed values for the
flow velocities $\bf u$, the
parallel force $F_\parallel $ and the magnetic field $\bf B$. Such a description is a good
approximation for the investigation of propagation of test particles. However, in many
interesting cases these quantities should be determined self-consistently, involving the
distributions $N_+$ and $N_-$.

We treat electrons as a massless fluid. In this case the Euler equation for electrons
is reduced to an expression for the electric field:

\begin{equation}
{\bf E +\delta E}= -\frac {\nabla P_e}{qn_e}-\frac 1c[{\bf u}_e\times ({\bf B +\delta B})],
\end{equation}
where ${\bf u}_e$ and $P_e$ are the velocity and the pressure of the electron fluid
respectively, $n_e$ is the electron number
density, $\bf \delta E$ and $\bf \delta B$ are the random components of the electric and the magnetic
fields respectively. The velocity ${\bf u}_e$ can be found from Maxwell's
equation

\begin{equation}
[{\bf \nabla \times (B+\delta B)}]=\frac {4\pi q}{c}
\int d^3p({\bf v}-{\bf u}_e)(f+\delta f),
\end{equation}
where $\delta f$ is the random component of the ion distribution,
and we have neglected the displacement current
by assuming a slow time evolution. It was also assumed that the plasma is pure
hydrogen and quasineutral.
Substitution of ${\bf u}_e$ from this equation into Eq. (16)  and averaging gives:

\[
{\bf E}= -\frac 1c[{\bf u}\times {\bf B}]-
\frac {1}{qn_i}\left( \frac 1{4\pi }[{\bf B\times [\nabla \times B}]]+
\nabla P_e+
\right.
\]
\begin{equation}
\left. \int d^3p{\bf p}\hat {St}f
\right).
\end{equation}
Here $n_i$ is the mean number density of ions and
we have used the formal definition of the scattering operator
$\hat {St}f=-\frac qc<\frac {\partial }{\partial {\bf p}}[{(\bf v}-{\bf u})
\times {\delta \bf B}]\delta f>$, where the
angle brackets mean the average over the random fluctuations of $\delta \bf B$ and
$\delta f$ of the magnetic
field and the momentum distribution. We also neglected the magnetic tension of the random
component. Generally
speaking, the scattering operator does not conserve momentum. Since the magnetic
inhomogeneities are frozen into the electron fluid, this additional force acts
on the electron fluid. As a result, an additional electric field due to the charge
separation appears. It is described by the last term in the parentheses of Eq. (18).
Such a field is indeed observed in hybrid simulations of collisionless shocks
(Quest \cite{quest88}).

We  multiply Eq.(1) containing the electric field (18) by ${\bf p}d^3p$ and integrate
over momentum space.
This gives the Euler equation of motion:

\begin{equation}
\rho \left( \frac {\partial {\bf u}}{\partial t}+
{\bf (u\nabla )u}\right) =
-\frac 1{4\pi }[{\bf B\times [\nabla \times B}]]-
\nabla (P_e+P_i).
\end{equation}
Here
\begin{equation}
P_i=\frac 13\int d^3p'v'p'(N_++N_-)
\end{equation}
is the pressure of ions and $\rho $ is the mean density.
The parallel electric field can be found from Eq.(18).

Since the Eq. (1) with the
electric field (18) conserves momentum, the same is true for Eqs. (9). Let us multiply
them by $p'$, integrate over $p'^2dp'$ and subtract the first from the second. This gives
after some algebra

\[
F_\parallel =\frac {1}{n_i}\left[ \frac 23\int d^3p'p'(\nu _++\nu _-)(N_+-N_-)+
{\bf b} \nabla P_i +\right.
\]
\begin{equation}
\left. {\bf b}\left( \frac {\partial {\bf u}}{\partial t}+
{\bf (u\nabla )u}\right) \int d^3p''
\left( \frac {p''}{v''}-\frac {p'}{v'}\right) (N_++N_-)
\right] .
\end{equation}
The last term term in the right hand side of this expression can be neglected, since the
force $F_\parallel $ is essential only for nonrelativistic particles which
determine the plasma density.

We shall assume an adiabatic evolution of the electron pressure
\begin{equation}
\frac {\partial P_e}{\partial t}+{\bf u}\nabla P_e+
\frac 53P_e\nabla {\bf u}=0,
\end{equation}
and frozen-in magnetic field.
The last term in Eq.(18) is essential for momentum conservation but can
be neglected in Faraday's induction equation:
\begin{equation}
\frac {\partial \bf B}{\partial t}=
[\nabla \times [{\bf u\times B}]].
\end{equation}
We have thus obtained the closed system of
Eqs. (12), (19), (22), (23) with expressions (10) for the frequencies $\nu _\pm$
and Eq. (21) for the parallel
force $F_\parallel $. In the next sections we apply these equations to the combined
problem of injection and nonlinear acceleration at a plane shock.

\section{Application for acceleration at the plane shock}

Let us consider the acceleration at the parallel one-dimensional shock. We can write the steady-state
version of Eq.
(9) for nonrelativistic particles in the shock frame as

\[
\left( u\pm \frac {v'}{2}\right) \frac {\partial N_\pm }{\partial z}+
\left( \pm \frac {F_\parallel }{2m}
-\frac {v'}3\frac {\partial u}{\partial z} \right) \frac {\partial N_\pm }{\partial v'}=
\]
\begin{equation}
\mp \nu _\mp (N_+-N_-).
\end{equation}
Here $m$ is the mass of particles and the parallel force
$F_\parallel =qE_\parallel -mu\partial u/\partial  z$.

The characteristics of this system of equations for the case $E_\parallel =0$
are given by

\begin{equation}
\left( \frac {v'}{u}\pm \frac 43+\frac {\sqrt{7}}{3}\right)
^{\frac 12\mp \frac {1}{\sqrt{7}}}
\left( \frac {v'}{u}\pm \frac 43-\frac {\sqrt{7}}{3}\right)
^{\frac 12\pm \frac {1}{\sqrt{7}}}
=\mathrm{const}
\end{equation}

They are shown in Fig.1. An upstream particle begins its motion with low
velocity from the upper left part of this
$u-v'$ plane and goes down the characteristics. The forward moving particles gain energy,
backward moving ones lose energy. At any moment the particle may change  hemisphere
and will continue the motion along another set of characteristics.
If the compression ratio of the shock is high enough, the forward moving particle can change
hemisphere in the vicinity of the
line $u=v'/2$ (shown by the {\it dash-dotted} line) and
return upstream along the backward characteristic with energy gain. It can again
change the hemisphere upstream and again move in a downstream direction etc. This
may be repeated many times and the accelerated particle goes to the right
beyond the part of $u-v'$ plane shown in Fig.1.

There is an another possibility for the initial acceleration. If the initial
speed of the backward moving particle upstream
is high enough, say 0.7$u_1$, it can reach the line $u=v'/2$ directly and return upstream with an
energy gain. However, the initial speed is rather high. The number of such particles
is small for high Mach number shocks.

\begin{figure}[t]
%\figbox*{}{}{
\includegraphics[width=6.0cm,angle=270]{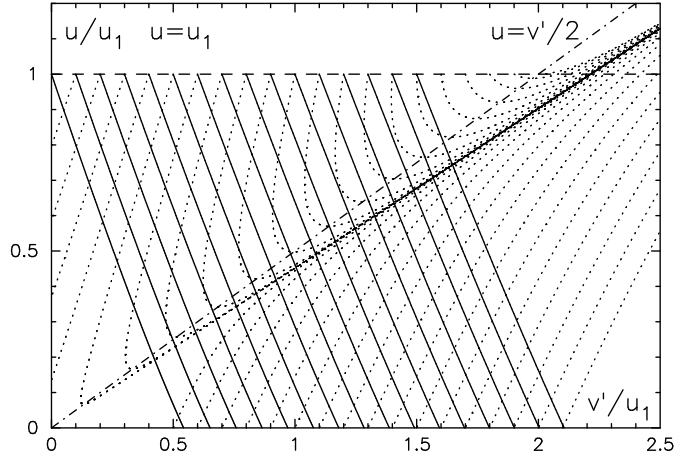}
\caption{Characteristics of the hyperbolic system of equations (24) for the
case $E_\parallel =0$. Characteristics for forward moving particles and
backward moving particles are
shown by {\it solid } and {\it dotted} lines respectively.}
\end{figure}

The characteristics of Eq. (24) depend on $F_\parallel $. The characteristics for
the case $F_\parallel =0$ are shown in Fig.2. This case is close to the real situation
because
the two first terms in Eq. (21) almost cancel each other (see also the numeric modeling below).
In other words, the electric force $qE_\parallel $ almost compensate the inertia
force $-mu\partial u/\partial z$ in the second term on the left-hand side of Eq. (24).
This is not a simple coincidence. The electric force is the consequence of the momentum
transfer and is approximately $-\partial P/\partial z/\rho $ that is
exactly $u\partial u/\partial z$.
The characteristics are determined by expressions

\begin{equation}
v'^4\pm \frac 83uv'^3=\mathrm{const} .
\end{equation}

\begin{figure}[t]
%\figbox*{}{}{
\includegraphics[width=6.0cm,angle=270]{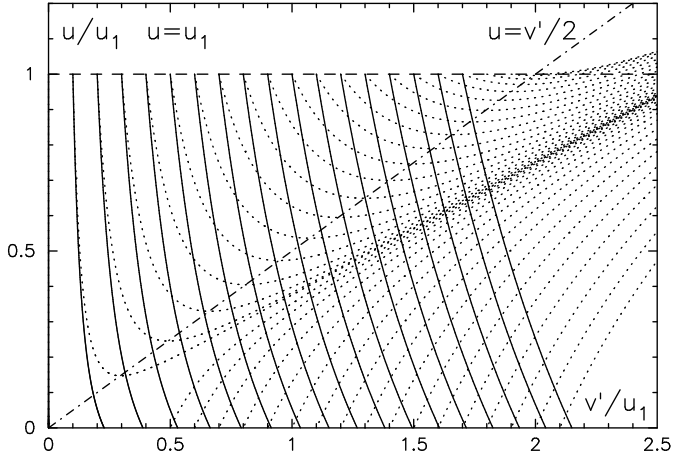}
\caption{Characteristics of the hyperbolic system of equations (24) for the
case $F_\parallel =0$. Characteristics for forward moving particles and
backward moving particles are
shown by {\it solid } and {\it dotted} lines respectively. }
\end{figure}

We find the analytical solution of Eq. (24) for the case when the
velocity profile is discontinuous: $u(z)=u_1$, $z<0$ and $u(z)=u_2$, $z>0$. In the
regions of constant velocity $u$ the system (24) can be rewritten as (to be
compared with  Gombosi et al. \cite{gombosi93})

\begin{equation}
u\frac {\partial N_+}{\partial z}=\frac {\partial }{\partial z}
\left( {v'^2}-4u^2\right) \frac {\lambda }{4v'}\frac {\partial N_+}{\partial z}
\end{equation}

\begin{equation}
N_-=N_++\left( 2{u}+v'\right) \frac {\lambda }{v'}\frac {\partial N_+}{\partial z}.
\end{equation}
It was also assumed that $E_\parallel =0$. The solution of these equations
upstream ($z<0$) is

\begin{equation}
N_\pm =N_\infty +(N_{1\pm }-N_\infty )
\exp \frac {4u_1v'z/\lambda }{{v'^2}-4{u_1^2}}.
\end{equation}

Here $N_\infty $ is the distribution
at $z=-\infty $, $N_{1+}$ and $N_{1-}$ are  the distributions $N_+$ and $N_-$
just upstream the shock. According to Eq. (28) they are related as

\[
(v'+2u_1)(N_{1+}-N_\infty )=(v'-2u_1)(N_{1-}-N_\infty ), \ v'>2u_1
\]
\begin{equation}
N_{1+}=N_{1-}=N_\infty ,\ v'<2u_1
\end{equation}

The solutions downstream are given by

\[
N_\pm =N_{2+}\left( \frac12+\frac {v'}{4u_2}\right)+
N_{2-}\left( \frac12-\frac {v'}{4u_2}\right)+
\]
\begin{equation}
\left( \frac {v'}{4u_2}\mp \frac12\right) (N_{2-}-N_{2+})
\exp \frac {4u_2v'z/\lambda }{{v'^2}-4{u_2^2}}.
\end{equation}
Here $N_{2+}$ and $N_{2-}$ are the distributions just downstream.
They are related as

\begin{equation}
N_{2+}=N_{2-}, \ v'>2u_2.
\end{equation}
This means that distributions $N_+$ and $N_-$ do not depend on $z$ for $v'>2u_2$.

We find the relation between the upstream and downstream
distributions $N_{1\pm }$ and $N_{2\pm }$. It depends on the solution of
Eq. (24) in the transition region. We consider the case
$F_\parallel =0$. Then the distributions upstream and downstream
are related by characteristics (26):

\begin{equation}
N_{1\pm }=F_\pm (v'^4\pm \frac 83u_1v'^3), \
N_{2\pm }=F_\pm (v'^4\pm \frac 83u_2v'^3).
\end{equation}
Here $F_\pm $ are two functions to be determined. Thus we have
the six equations (30), (32), (33) for the six unknown functions
$N_{1\pm }$, $N_{2\pm }$ and $F_\pm $.

Let us consider the case $N_\infty =\delta (v'-v_0)/(2v_0^2)$. Since the change of
velocity in the transition region is governed by the characteristics (26), the
solution is the following sum of $\delta $-functions:

\begin{equation}
N_{1\pm }=A_{0}\delta (v'-v_{0})/(v_{0})^2 +
\sum ^2_{j=1}\sum ^\infty _{i=1}A^j_{i\pm }\delta (v'-v^j_{i})/(v^j_{i})^2,
\end{equation}
\begin{equation}
N_{2\pm }=\sum ^2_{j=1}\sum ^\infty _{i=0}B^j_{i\pm }\delta (v'-w^j_{i})/(w^j_{i})^2.
\end{equation}
Here $A^j_{i\pm }$, $B^j_{i\pm }$, $v^j_{i}$ and $w^j_{i}$ are sequences to be determined.
The index $j$ may have the values 1 and 2. These values correspond to the two possibilities
of injection of forward and backward moving particles. They were described after Eq. (25). The
different values of the index $i$ correspond to consequent states of the one particle moving
between upstream and downstream regions of the shock.

If the upstream sequences $A^j_{i\pm }$ are known for $v'>2u_1$
the downstream  ones can be found from characteristics (26). Using the properties of
$\delta $-functions we find that they are given by

\begin{equation}
B^j_{i\pm }= \frac {v^j_i+2u_1}{w^j_i+2u_2}A_{i+}, \ w^j_i>2u_2
\end{equation}
and the downstream velocity $w^j_i$ is the solution of equation

\begin{equation}
(v^j_i)^4+ \frac 83u_1(v^j_i)^3=(w^j_i)^4+\frac 83u_2(w^j_i)^3.
\end{equation}

The next $A^j_{i+1\pm }$ are

\begin{equation}
A^j_{i+1-}= \frac {w^j_i-2u_2}{v^j_{i+1}-2u_1}B^j_{i-}, \
A^j_{i+1+}=B^j_{i-}\frac {w^j_{i}-2u_2}{v^j_{i+1}+2u_1}
\end{equation}
where the velocity $v^j_{i+1}$ is the solution of the equation

\begin{equation}
(v^j_{i+1})^4- \frac 83u_1(v^j_{i+1})^3=(w^j_i)^4-\frac 83u_2(w^j_i)^3.
\end{equation}

Eqs. (36)-(39) can be used recurrently to calculate the
sequences $A^j_{i\pm }$, $B^j_{i\pm }$, $v^j_{i}$ and $w^j_{i}$.

Let us introduce the critical velocities $v_{c\pm }$. They are the
solutions of equations:

\begin{equation}
v^3_{c\pm }(3v_{c\pm }\pm 8u_1)=(48\pm 64)u^4_2.
\end{equation}

The initial velocity $v_0$ and $A_{0}=1/2$ are given. There are four
cases depending on the initial velocity $v_0$ and velocities $v_{c\pm }$.

1) $v_0<v_{c-}$. Forward and backward moving particles reach the downstream region with
velocities smaller than 2$u_2$ and are further advected. The injection in acceleration
does not occur. The downstream coefficients $B^j_{0\pm }$ are given by
\[
B^1_{0+}=A_0\frac {v_0+2u_1}{w^1_0+2u_2}, \ B^2_{0-}=0
\]
\begin{equation}
B^2_{0-}=A_0\frac {v_0-2u_1}{w^2_0-2u_2}, \ B^1_{0-}=0
\end {equation}
where the velocities $w^1_0$ and $w^2_0$ are the solutions of equations:
\[
(v_{0})^4+\frac 83u_1(v_{0})^3=(w^1_0)^4+\frac 83u_2(w^1_0)^3
\]
\begin{equation}
(v_{0})^4-\frac 83u_1(v_{0})^3=(w^2_0)^4-\frac 83u_2(w^2_0)^3.
\end {equation}
All other coefficients $A^j_{i\pm }=0$, $B^j_{i\pm }=0$ for $i>0$.

2)$v_{c-}<v_0<v_{c+}$. The most important case for injection. The forward moving particle
reaches the downstream region with a velocity less than $2u_2$ and again is not injected into
acceleration. The coefficients $B^1_{i\pm }$ and $A^1_{i\pm }$ are the same as in the previous case.

The backward moving particle goes
along characteristics and returns upstream (see Fig.2). Using properties of $\delta $-functions
in Eq. (33) and conditions (30) we found

\begin{equation}
A^2_{1-}=A_{0}\left| \frac {v_0-2u_1}{v^2_1-2u_1}\right| ,
\ A^2_{1+}=A^2_{1-}\frac {v^2_1-2u_1}{v^2_1+2u_1}.
\end{equation}
Here the velocity $v^2_1$ is the solution of the equation

\begin{equation}
(v_0)^4- \frac 83u_1(v_0)^3=(v^2_1)^4-\frac 83u_1(v^2_1)^3.
\end{equation}
It is clear that $v^2_1$ is close to $8u_1/3$ (see Fig.2).
The other coefficients $A^2_{i\pm }$ and $B^2_{i\pm }$ can be found consecutively using Eqs. (36)-(39).

3) $v_{c+}<v_0<2u_1$. Forward and backward moving particles reach the downstream region with
velocities greater than $2u_2$ and are injected into acceleration. The coefficients $A^2_{i\pm }$ and
$B^2_{i\pm }$ are the same as in the previous case. The coefficients $B^1_{0\pm }$ are given by

\begin{equation}
B^1_{0\pm }= \frac {v_0+2u_1}{w^1_0+2u_2}A_{0},
\end{equation}
and the downstream velocity $w^1_0$ is the solution of the equation

\begin{equation}
(v_0)^4+ \frac 83u_1(v_0)^3=(w^1_0)^4+\frac 83u_2(w^1_0)^3.
\end{equation}
The coefficients $A^1_{1\pm }$ can be found from equations Eqs. (38) and (39). Now the
coefficients $B^1_{i\pm }$ at $i>0$ and $A^1_{i\pm }$ at $i>1$ can be found consecutively using Eqs. (36)-(39).

4) $v_0>2u_1$. The coefficients $A^j_{i\pm }$ and $B^j_{i\pm }$ at $i>0$
can be found consecutively using Eqs. (36)-(39).

Using Eqs. (36) and (38) one can relate $A^j_{i+1+}$ and $A^j_{i+}$:
\begin{equation}
A^j_{i+1+}=A^j_{i+}\frac {w^j_{i}-2u_2}{v^j_{i+1}+2u_1}\frac {v^j_i+2u_1}{w^j_i+2u_2}
\end{equation}

\begin{figure}[t]
%\figbox*{}{}{
\includegraphics[width=6.0cm,angle=270]{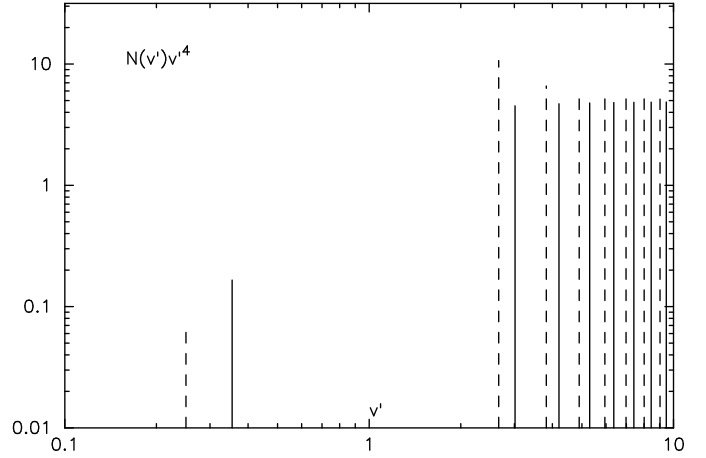}
\caption{The analytic solution for $u_1/u_2=4$ and $v_0=0.25u_1$ for the
case $F_\parallel =0$. The downstream and upstream distributions are
shown by {\it solid } and {\it dashed} lines respectively. The height of the
vertical lines corresponds to the coefficients in Eqs. (34) and (35)}
\end{figure}

It is easy to show that for large $w^j_i$ and $v^j_i$ the change of
velocity

\begin{equation}
w^j_i-v^j_i\approx v^j_{i+1}-w^j_i\approx \frac 23(u_1-u_2).
\end{equation}

Using Eq. (47) we find that for high velocities

\begin{equation}
A^j_{i+}/(v^j_i)^2\sim (v^j_i)^{-3u_1/(u_1-u_2)}, \ v^j_i>>u_1,u_2
\end{equation}
that is exactly the spectrum of particles produced by diffusive shock acceleration.

The solution for $u_1/u_2=4$ and $v_0=0.25u_1$ is shown in Fig.3. The
critical velocities are $v_{c-}\approx 0.2u_1$ and $v_{c+}\approx 0.35u_1$ in this case.
The far upstream distribution is shown by the dashed line on the left. The forward moving
particles of this distribution reach the downstream region with slightly higher velocity
(solid line on the left). The backward moving particles are directly accelerated along
the characteristic and return upstream with a velocity close to $8u_1/3$ (first dashed line
on the right). Particles are accelerated further, moving between downstream and upstream
regions of the shock (the other lines on the right). The similar $\delta $-function
spectrum was obtained by Bogdan and Webb (\cite{bogdan87}) for cosmic ray acceleration in
the so-called two-streaming
approximation of Fisk and Axford (\cite{fisk69}).

The spectrum will be smoother if one takes into account the scattering of particles in the
transition region.

\section{Numeric modeling of collisionless shocks}

\begin{figure}[t]
%\figbox*{}{}
\includegraphics*[width=9.0cm]{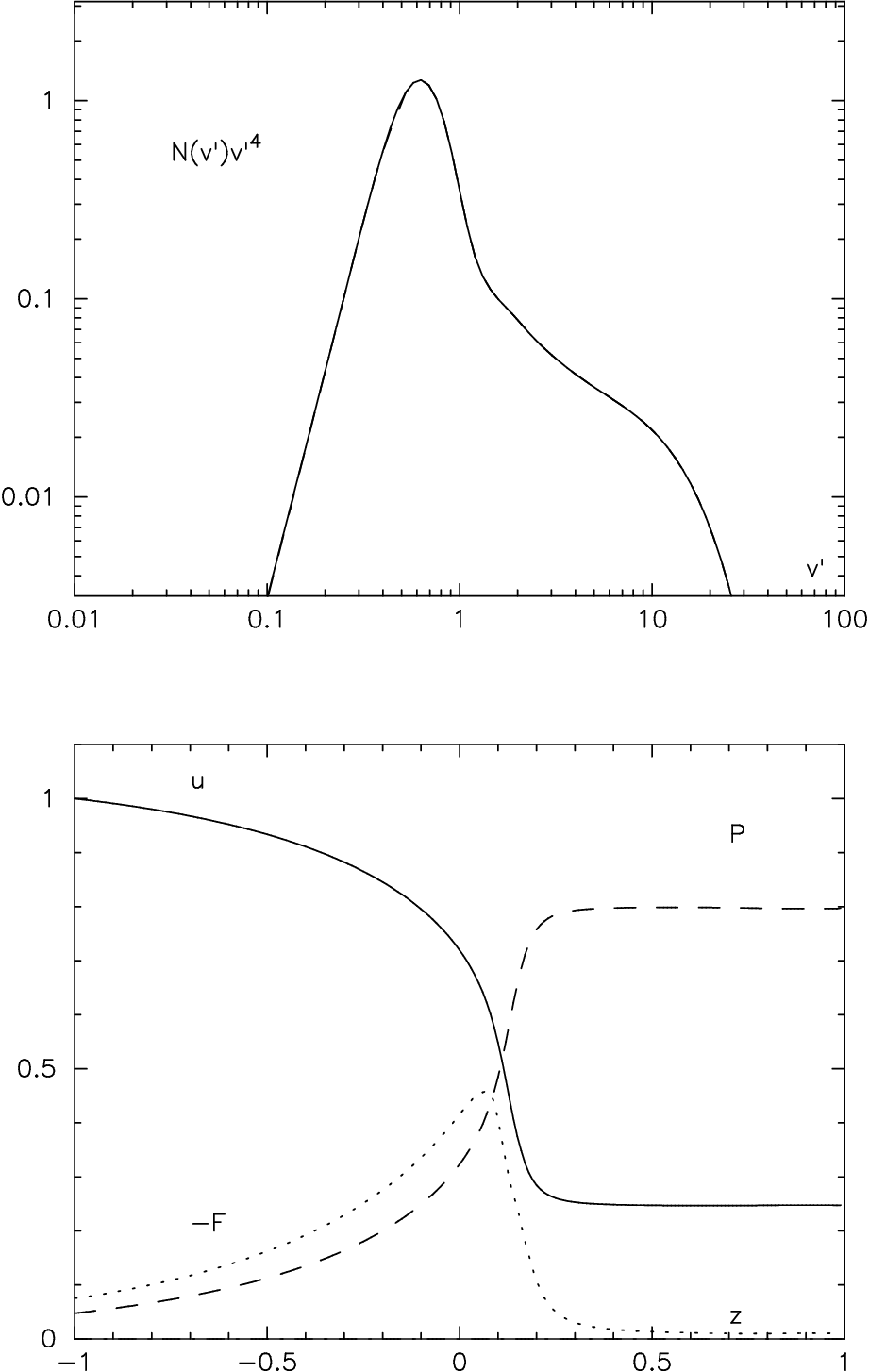}
\caption{The results of simulation of the parallel shock with Mach number 3.77,
$T_e/T_i=0$ and energy independent free path $\lambda =0.2$.
The forward and backward particle distributions downstream $N_+$ and $N_-$ are
shown on the top panel by {\it solid } and {\it dashed} lines respectively.
Plasma velocity $u$, pressure of ions $P$ and electric force $F=qE_\parallel $
are shown on the bottom
 panel by {\it solid }, {\it dashed} and {\it dotted} lines respectively.
The total compression ratio $r=4.05$ is obtained.}
\end{figure}

We use the system of equations (12),(19),(22),(23) for the modeling of
steady-state one-dimensional nonrelativistic shocks.
We neglect the dynamical effects of the magnetic
field, which means  that the magnetic field produces only kinematic effects that are
essential for oblique shocks. The upstream plasma state is then determined only by the
sonic shock Mach number $M$ and by the ratio of electron and ion temperatures $T_e/T_i$.

\begin{figure}[t]
%\figbox*{}{}
\includegraphics*[width=9.0cm]{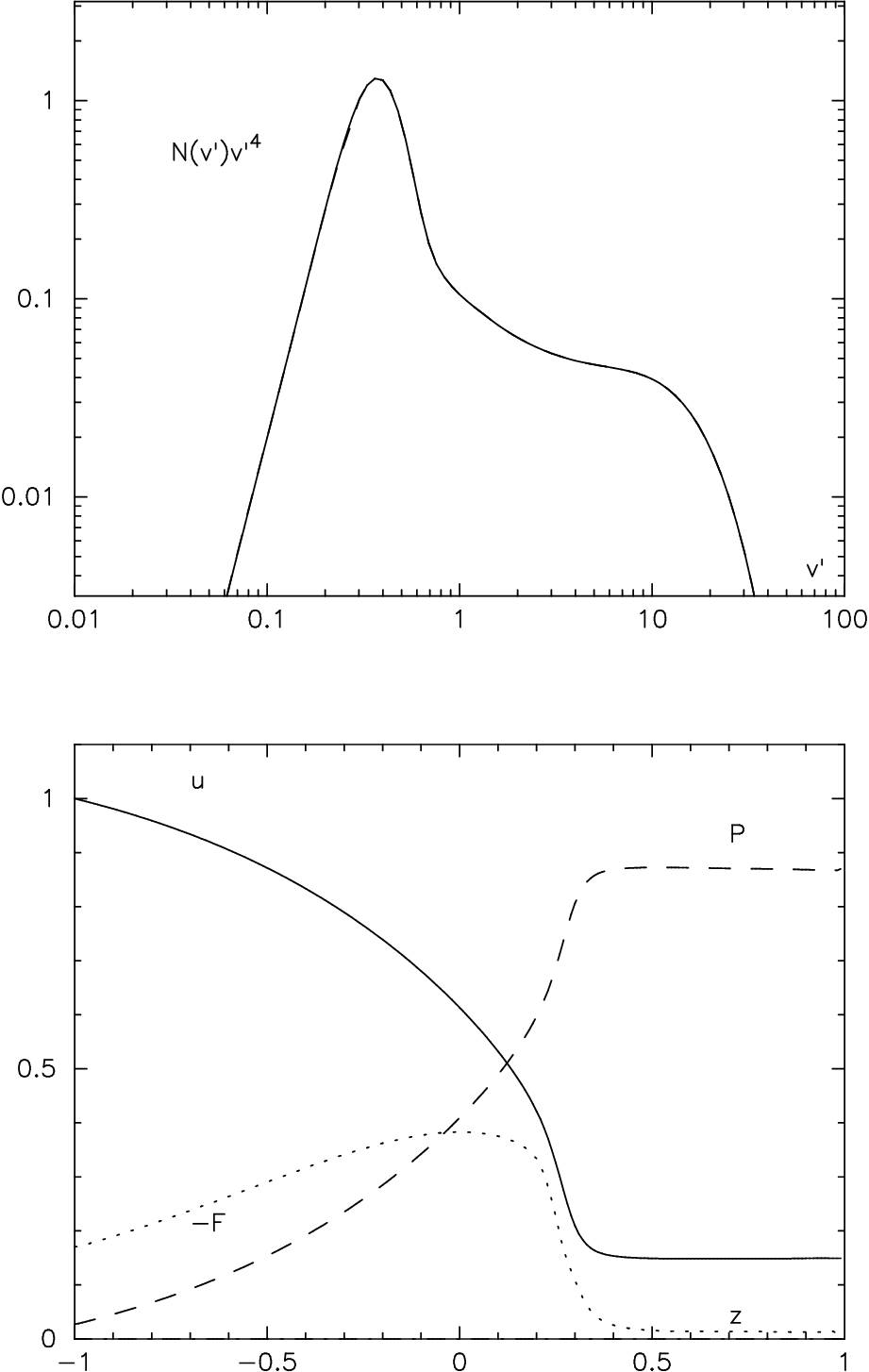}
\caption{The results of simulation of the parallel shock with Mach number 7.75,
$T_e/T_i=0$ and energy independent free path $\lambda =0.2$.
The forward and backward particle distributions downstream $N_+$ and $N_-$ are
shown on the top panel by {\it solid } and {\it dashed} lines respectively.
Plasma velocity $u$, pressure of ions $P$ and electric force $F=qE_\parallel $
are shown on the bottom panel by {\it solid }, {\it dashed} and {\it dotted} lines respectively.
The total compression ratio $r=6.67$ is obtained.}
\end{figure}

A so-called Alfv\'{e}n heating is not included in our model. It is well known
that the energetic particles upstream of the shock can effectively generate
Alfv\'{e}n waves due to streaming instability (Wentzel \cite{wentzel69},
Bell \cite{bell78}). The damping of these waves results in the heating of the shock
precursor (V\"{o}lk \& McKenzie \cite{voelk82}, McKenzie \& V\"{o}lk \cite{mckenzie82}).
This heating is essential,
since it decreases the effective Mach number and compression ratio of the shock.
As shown in the Appendix, this effect can be formally included using the
relation between sonic and Alfv\'{e}n Mach numbers of the shock:

\begin{equation}
M^2=\frac {M_s^2M_a}{M_a+\frac {20}{39}M_s^2}
\end{equation}
The results obtained below for shocks with sonic Mach number $M$ without Alfv\'{en}
heating can be formally
used for shocks with Alfv\'{e}n heating and with Alfv\'{e}n and sonic Mach numbers $M_a$ and
$M_s$ respectively.

The details of the numeric method are given in the Appendix. The plasma flow with
unity velocity and  unity density enters the simulation box from its left boundary. The
pressure at the right boundary was adjusted according to the motion of the shock in order
to reach the steady state. We used the free escape boundary condition for the particle
distribution at the left boundary.

We prescribe the dependence of the free path of particles on the velocity $v'$ and
space coordinate $z$. We used the  free path $\lambda $ that is independent on $z$. This is not
a strong limitation. Indeed, in the rather general case of separable dependence of $\lambda $
on $v'$ and $z$, the results obtained depend only on the integral $\int dz/\lambda (z)$. Thus
our results can be used also for this more general case.

The energy dependence of the mean free path $\lambda $ is determined by the spectrum of magnetic
inhomogeneities and by the mechanism of scattering in the vicinity of a pitch angle of 90 degrees. Since
the low frequency waves propagating along the magnetic field are subject to nonlinear
wave steepening,
one can expect that the magnetic spectra are close to $k^{-2}$ and the corresponding free path
does not depend on energy. Such spectra (or slightly flatter ones) were indeed observed
in the hybrid plasma simulations of parallel collisionless shocks
(Giacalone et al. \cite{giacalone93}, Giacalone \cite{giacalone04}) and in the vicinity of
the Earth bow shock in the solar wind. In the  results presented here we used the energy independent
free path.

We limit ourselves to the case of cold electrons $T_e/T_i=0$, as
the dissipation of the Alfv\'en waves in the precursor may result in the preferable heating
of ions. The solar corona is an example of this.

We start with the case of parallel shocks with
Mach numbers 3.87 and 7.75. According to Eq. (50) the corresponding Alfv\'{e}n Mach numbers
are 7.5 and 30 that can be applied to the interplanetary and supernova shocks
respectively. The downstream particle distributions are shown in the top
panels of Fig.4 and 5. The velocity and gas pressure profiles are shown in the bottom
panels.

The simulated shocks demonstrate clear sub-shocks with a width of about one half
of the free path $\lambda $ and a smooth precursor produced by the particles in the
high energy tail that is seen in the top panels of Fig.4 and 5. It was found that the main
part of the downstream distribution in Fig.4 and 5 is the adiabatically compressed far upstream
distribution which was taken to be Maxwellian. The total electric potential corresponding
to the electric force $qE_\parallel $ is about 0.2-0.3 $mu_1^2$ according to Fig.4 and 5.

The total
compression ratio of the shock can be found from the following simple estimate.

In the precursor region the thermal particles are heated adiabatically. When their
velocities become comparable to the  plasma velocity, the backward moving particles
are accelerated more efficiently (see Fig.2) and injected for further
acceleration. The compression ratio should be large enough for this to be the case.
For estimation we assume that distribution of upstream particles is
proportional to $\delta (v'-v_{\mathrm{T}})$,
where $v_\mathrm{T}=\sqrt{3/5}u_1/M$ is the thermal
velocity of the plasma. The maximum shock downstream velocity
 $u_2$ that is necessary for
injection of such particles according to Fig.2 and characteristics (26)
is

\begin{equation}
u_2=\frac 12v_\mathrm{T}\left( 8\frac {u_1}{v_\mathrm{T}}-3\right) ^{1/4} .
\end{equation}
Thus the thermal velocity $v_T$ should be larger than the critical
velocity $v_{c-}$ (see Sect. 3).
The second term in parentheses can be neglected. Writing the thermal
velocity $v_\mathrm{T}$ in terms of Mach number $M$ we obtain:
\begin{equation}
r=2^{1/4}\left( \sqrt{5/3}M\right) ^{3/4}\approx 1.44M^{3/4}.
\end{equation}
This formula is in good agreement with the simulated compression ratio . It was
found that it is valid also for smaller free paths, when the maximum energy of the high
energy tail is larger.

\begin{figure}[t]
%\figbox*{}{}
\includegraphics*[width=9.0cm]{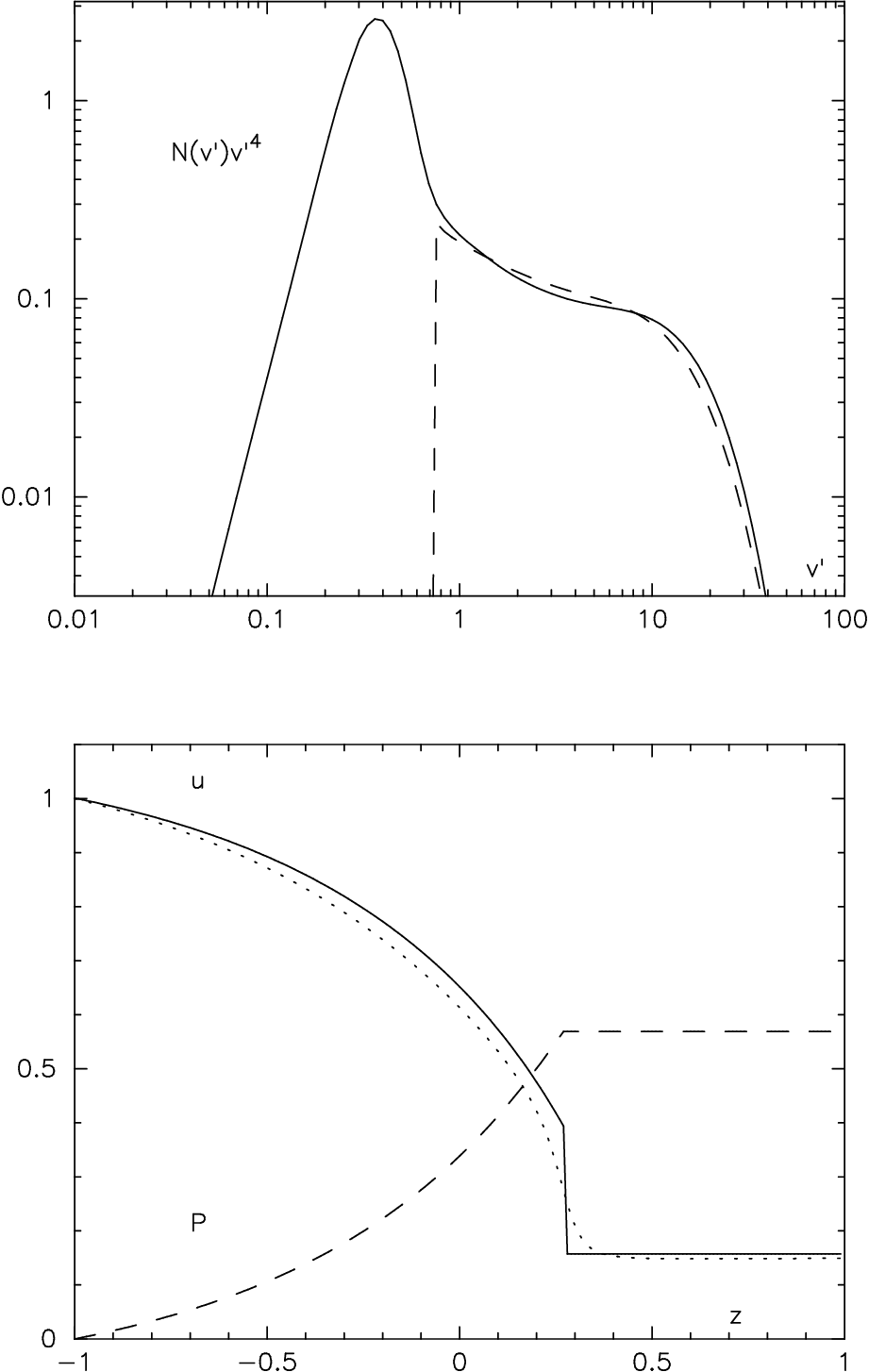}
\caption{Comparison of our model with
the cosmic ray modified shock approach for a parallel shock with Mach number 7.75,
$T_e/T_i=0$ and energy independent free path $\lambda =0.2$.
The downstream distributions for our model and for the  cosmic ray approach are
shown in the top panel by {\it solid } and {\it dashed} lines respectively.
Plasma velocity $u$ from our simulations, pressure of ions $P$ and velocity
profile for the cosmic ray approach are
shown in the bottom panel by {\it dotted}, {\it dashed} and {\it solid } lines respectively.}
\end{figure}

This compression ratio is close to the maximum
possible compression ratio of the idealized system that includes the infinitely
thin gas sub-shock and the viscous precursor produced by energetic particles. Using
the Ranke-Hugoniot conditions at the thermal sub-shock one can obtain for this case:

\begin{equation}
r=2.5\left( M^2/5 \right) ^{3/8}\approx 1.37M^{3/4}.
\end{equation}
This number is also close to the value found in numerical simulations of
shock waves, modified by the cosmic ray pressure
(see e.g. Berezhko \& Ellison \cite{berezhko99}). The compression ratio of
the thermal sub-shock is 2.5.

The comparison of our results and results obtained for this simplified
description are shown in Fig.6.
The Eq. (14) was solved in the upstream region together with the Euler equations.
The top panel shows the comparison of
downstream spectra obtained in these approaches. The velocity profiles are
compared in the bottom panel. We find a very good agreement of these two
 methods.
We estimate the  injection efficiency at injection velocity
$v_{\mathrm{inj}}=2u_{\mathrm{sub}}$
to be about 0.024. Here $u_{\mathrm{sub}}$ is the sub-shock velocity.
This number is in reasonable agreement with results of hybrid simulations
(cf. Giacalone et al. \cite{giacalone93}, Ellison et al. \cite{ellison93})
and the Earth bow shock observations (cf. Ellison et al. \cite{ellison90}).

We have also performed modeling of a quasi-parallel shock. The results obtained for
a Mach number of 7.75 and an angle between the shock normal and magnetic field of
$\theta =15^{\mathrm{o}}$ are
shown in Fig.7.

\begin{figure}[t]
%\figbox*{}{}
\includegraphics*[width=9.0cm]{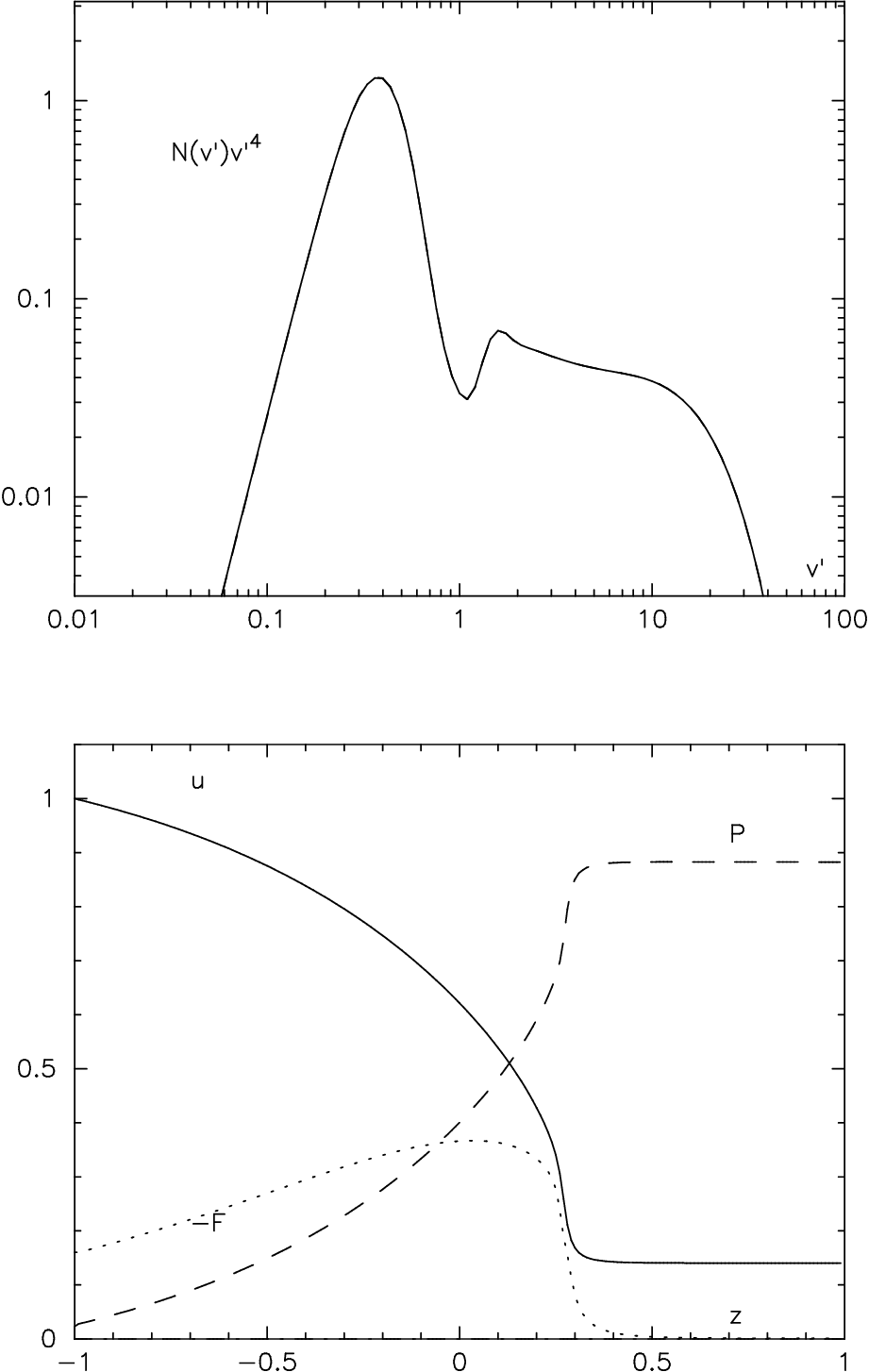}
\caption{Simulation of the quasi-parallel shock with $\theta =15^{\mathrm{o}}$,
Mach number M=7.75,
$T_e/T_i=0$ and energy independent free path $\lambda =0.2$.
The forward and backward particle distributions $N_+$ and $N_-$ are
shown in the top panel by {\it solid } and {\it dashed} lines respectively.
Plasma velocity $u$, pressure of ions $P$ and electric force $F=qE_\parallel $
shown in the bottom panel by {\it solid }, {\it dashed} and {\it dotted} lines respectively.
The total compression ratio $r=7.21$ is obtained.}
\end{figure}

For such oblique shocks the dip appears between the main part of the thermal distribution
and the high-energy tail (see the top panel). The injection efficiency is the same as in the
previous case.
$15^{\mathrm{o}}$ is the maximal value of $\theta $ that allows us to model steady state shocks with
$M=7.75$.
Even this value is rather large, because it corresponds to the normal angle 40$^{\mathrm{o}}$ upstream
of the thermal subshock.
For larger $\theta $ this dip becomes more pronounced and the thermal subshock
width drops to a grid step. In some cases some instability in the downstream region
was observed.
It is possible to model the larger values of $\theta $ for the smaller Mach numbers
(e.g. $\theta =30^{\mathrm{o}}$ for $M=3.87$).
Thus our approach can be used only for quasiparallel shocks with the normal
angle less than 40$^{\mathrm{o}}$ just upstream of the thermal subshock.

If the magnetic field is oblique enough, it prevents particles
from returning from downstream along magnetic lines and the
heating of particles is impossible. Since it is necessary to have
the energetic particles downstream for the existence of a shock,
we expect that the transition region will be squeezed to widths of
the order of the ion gyroradius. Heating then may take place in a
different regime, where the drift motion of particles
perpendicular to magnetic field is essential. The injection
efficiency also may be very different.

\section{Conclusion}

Many astrophysical problems deal with collisionless plasma. Different low
frequency instabilities may produce magnetic inhomogeneities that can
scatter particles and play the  role of thermal collisions. When unstable
waves propagate preferably along the mean magnetic field, the resonant
scattering of particles is weak in the vicinity of 90 degrees pitch angle. In this
case it is possible that particle distributions are almost isotropic in
both hemispheres.

Following the approach of Isenberg (\cite{isenberg97}) we
derive the general transport equation (12) (or equivalent Eq.(9))
which takes into account the
motion of the medium and energy changes of particles. Our consideration is
not limited by high  particle velocities and
by small anisotropies of particle distribution.

In the case of high particle velocities and small anisotropies the
equation derived can be reduced to the standard cosmic ray
transport equation. In the case of negligible energy changes and advective
transport the equation
may be transformed to the so-called telegraph equation.

The equations derived can be solved together with magnetohydrodynamic equations (see Sect.3).
We used the fluid approximation for the electron transport. This is justified
if electrons are effectively scattered by magnetic inhomogeneities. If this is
not the case, one can use the equations (12) for electron transport. In this way
electron injection at collisionless shocks can be investigated.
%We leave this
%opportunity for future investigations.

The transport equations (12) with Eq. (19), (22) and (23) can be used
for the solution of different astrophysical problems when the self-consistent
determination of the flow velocity, magnetic field etc. is necessary.

We apply the equation derived to investigate particle acceleration
and injections at astrophysical shocks. For the prescribed velocity profile
we have found the exact analytical solution (see Sect. 4).

We also have performed the modeling of collisionless shocks and found a
 formula for the shock compression ratio (see Eq.(52)).
This ratio is
very close to the maximal possible value in the framework of
the simplified approach used for cosmic ray modified shocks.
The corresponding thermal sub-shock ratio is 2.5.

We found that this feature is universal and does not depend on the maximum
energy of the accelerated particles. This was checked in our simulations up to
the maximum momentum $p_{\max }\sim 100 \ mu_1$ which was limited by the step of
our uniform spatial grid. This property will not change for
higher maximum energies corresponding to the supernova environment. On the other
hand we found that the injection efficiency $\eta $ is not universal.
For $p_{\max }\sim 10\ mu_1$ it is about 0.024 for an  injection
velocity equal to two sub-shock velocities. It slowly decreased as the maximum
energy increases and is about 0.003 for young supernova remnants.
This is because at small maximum energies the thermal ions "feel" a larger
compression ratio than the thermal subshock compression ratio since the subshock
and precursor width are not strongly different.

We found that the equations derived can be used to investigate injection and
acceleration in quasiparallel shocks with the subshock normal angle $\theta <40^{\mathrm{o}}$. The
dissipation at more oblique shocks is described by a different mechanism which
takes into account the motion of particles in the direction perpendicular to the magnetic
field. The injection efficiency at such shocks is unknown, but it is possible
that it is rather low. We confirm the previous findings that quasiparallel
shocks are very efficient accelerators with high injection efficiency
(see e.g Ellison \& Eichler \cite{ellison84}, Giacalone et al. \cite{giacalone93},
Malkov \& Drury \cite{malkov01})
which does not
depend on the angle $\theta $ (Giacalone et al. \cite{giacalone97}).

% (i) one column figure, will be floated to top of next column
%
%\begin{figure}[t]
% \vspace*{2.0mm} % just in case for shifting the figure slightly down
%\includegraphics[width=8.3cm]{outfile.ps} % .eps for Latex,
%                                            % pdfLatex allows .pdf, .jpg, .png and .tif
%\caption{Figure caption text.}
%\end{figure}

% (iii) 1 1/2 column figure with caption on the side, will be floated
% to top of next page
% \begin{figure*}[t]
% \figbox*{}{}{\includegraphics*[width=11.0cm]{figfile.eps}}
% \caption{Figure caption text.}
% \end{figure*}
%\section{Balancing}
%The columns of the last page can be balanced either by using the
%command \verb/\balance/ somewhere in the first column of the last page
%or by explicitely put \verb/\vadjust{\newpage}/ at the correct place.
% without the \verb/ / command
\begin{acknowledgements}
The author is grateful to Heinz V\"olk for fruitful discussions of
the collisionless shock physics. The author also thanks the  anonymous referee for
important comments.
\end{acknowledgements}
\appendix
\section{Formal inclusion of  Alfv\'{e}n heating}

The heating of the cosmic ray precursor is described by the
following equations (V\"{o}lk \& McKenzie \cite{voelk82},
McKenzie \& V\"{o}lk \cite{mckenzie82}):
\begin{equation}
\frac {\partial }{\partial z}u\rho =0
\end{equation}
\begin{equation}
\rho u\frac {\partial u}{\partial z}=-\frac {\partial P_g}{\partial z}
-\frac {\partial P_c}{\partial z},
\end{equation}

\begin{equation}
u\frac {\partial P_g}{\partial z}+\gamma _gP_g\frac {\partial u}{\partial z}=
(\gamma _g-1)V_a\frac {\partial P_c}{\partial z}
\end{equation}
Here $V_a$ is the component of Alfv\'{e}n velocity parallel to the shock
normal, $P_g$ and $P_c$ are the pressure of the gas and cosmic rays respectively,
and $\gamma _g$ is the gas adiabatic index.
The third equation describes the gas heating due to the damping of Alfv\'{e}n waves
generated by the cosmic ray streaming instability. The second equation is the Euler
equation of motion.

The gradient of the cosmic ray pressure can be found from Eq. (A.2) and substituted
into Eq. (A.3). For high Mach number shocks the  solution can be written as
\[
P_g=\left( P_{g1}+\rho _1u_1V_{a1}\frac {\gamma _g-1}{\gamma _g+\frac 12} \right)
\left( \frac {u_1}{u}\right) ^{\gamma _g}-
\]
\begin{equation}
\rho _1u_1V_{a1}\frac {\gamma _g-1}{\gamma _g+\frac 12}
\left( \frac {u}{u_1}\right) ^{1/2}.
\end{equation}
Here $P_{g1}$ and $V_{a1}$ are the gas pressure and Alfv\'{e}n velocity
in the medium, in which the shock propagates with speed $u_1$.

For high Mach number shocks the Alfv\'{e}n heating is essential only at the
very beginning of the precursor. In the rest, the gas is heated adiabatically
and the last term in Eq. (A.4) can be neglected. The gas pressure $P_{g1}$
and Alfv\'{e}n velocity $V_{a1}$ can be expressed in terms of Alfv\'{e}n and
sonic Mach numbers $M_a$ and $M_s$ respectively. As a result, we obtain the
sonic Mach number $M$ of the shock without Alfv\'{e}n heating, which is equivalent to
the shock with the Alfv\'{e}n heating:
\begin{equation}
M^2=\frac {M_s^2M_a}{M_a+\gamma _g\frac {\gamma _g-1}{\gamma _g+\frac 12}M_s^2}
\end{equation}
For $\gamma _g=5/3$ we obtain the formula (50) from the main part.

\section{Numeric method}

We used the implicit finite difference scheme for the solution of the one-dimensional
version of Eq. (12):
\begin{equation}
\frac {\partial n}{\partial t}=L_z n+L_vn.
\end{equation}
Here $n$ describes the pair of forward and backward particle distributions
$n_\pm =v'^3N_\pm $, and
$L_z$ and $L_v$ are the operators in the coordinate and velocity space
respectively:
\begin{equation}
L_zn=-\frac {\partial }{\partial z}\left( u\pm \frac 12v'b_z\right) n_\pm,
\end{equation}
\begin{equation}
L_vn=-v'\frac {\partial }{\partial v'}\left( \pm \frac {F_\parallel }{v'}-
\frac{1}{3}\frac {\partial u}{\partial z}\right) n_\pm +
\nu _\mp n_\mp -\nu _\pm n_\pm.
\end{equation}
The finite difference version of Eq. (B.1) was solved in two steps (Godunov \cite{godunov71}):

\begin{equation}
\frac {n_{k+1/2}^{i,j}-n_k^{i,j}}{\tau }=L^{dif}_vn_{k+1/2}^{i,j}+L^{dif}_{z}n_{k}^{i,j},
\end{equation}
\begin{equation}
\frac {n_{k+1}^{i,j}-n_{k+1/2}^{i,j}}{\tau }=L^{dif}_zn_{k+1}^{i,j}-L^{dif}_{z}n_{k}^{i,j}.
\end{equation}
Here $n_k^{i,j}$ is the value of distribution $n(z,v',t)$ at $z=i\Delta z$, $v'=v_j$,
$t=k\tau $, where $\tau $ is the time step, $\Delta z$ is the grid size in
$z$-direction and $v_j$ is the value of velocity $v'$ at the knot of the velocity grid
with number $j$.
We use the following finite difference analog $L^{dif}_z$ of the operator
$L_zn=-\partial /\partial zAn$ (Kota et al. \cite{kota82}):
\[
L^{dif}_zn_k^{i,j}=\left( \frac 16A_k^{i+2}n_k^{i+2,j}-A_k^{i+1}n_k^{i+1,j}+
\frac 12A_k^{i}n_k^{i,j}+
\right.
\]
\begin{equation}
\left. \frac 13A_k^{i-1}n_k^{i-1,j}\right) /{\Delta z}, \ A_k^i<0,
\end{equation}
\[
L^{dif}_zn_k^{i,j}=\left( -\frac 16A_k^{i-2}n_k^{i-2,j}+A_k^{i-1,j}n_k^{i-1,j}-
\frac 12A_k^{i}n_k^{i,j}-
\right.
\]
\begin{equation}
\left. \frac 13A_k^{i+1}n_k^{i+1,j}\right) /{\Delta z}, \ A_k^i>0.
\end{equation}
A similar operator was used for the finite difference
analog $L_v^{dif}$ of the velocity operator (B.3).

The numeric scheme (B.4),(B.5) with these operators
has a third order accuracy on $z$ and $v'$ and first
order accuracy on $t$.

The parallel force $F_\parallel $ was recalculated at each time step according to
Eq. (21). The medium velocity $u_k^{i}$ at each time step was calculated
according to the finite difference version of the Euler equation (19):

\[
\frac {\rho _{k+1}^iu_{k+1}^i-\rho _{k}^iu_{k}^i}{\tau }= -\frac
{\rho _{k+1}^{i}(u_{k+1}^{i})^2-\rho _{k+1}^{i-1}(u_{k+1}^{i-1})^2}{\Delta z}+
\]
\begin{equation}
\frac {P_{k+1}^{i-1}-P_{k+1}^{i+1}}{2\Delta z}.
\end{equation}
Here $P_k^i$ is the sum of the ion and electron pressures. The ion pressure and plasma
density were calculated from
the particle distribution at the each time step.
The electron pressure was found from the adiabatic equation
of state. The Eq. (B.8) is the quadratic equation for the velocity $u_{k+1}^i$. The gas
pressure at the right boundary $P_k^b$ was prescribed in accordance with the shock motion:
\begin{equation}
P_k^b=P_{k-1}^b+\frac 12(Z_k^{s}-Z_{k-1}^s).
\end{equation}
Here $Z_k^s$ is the position of the shock. When the
shock moves to the right boundary, the pressure
increases. This permits us to reach steady state.

We use the hyperbolic tangent for the initial velocity profile. The initial
particle distribution was Maxwellian with a temperature and density dependence
corresponding to the momentum and mass conservation. At times of about 60-100 in dimensionless
units the system reaches the steady state. We use 200 grid points in the $z$-direction,
100 grid points for the logarithmic $v'$ grid and the time step between $0.005$ and $0.1$ for
different runs. We obtained the total momentum, energy and mass conservation with
several percent accuracy.

\end{document}